\documentclass[twocolumn,aps,prl,floatfix,letterpaper,
                 superscriptaddress,showpacs]{revtex4}

\usepackage{amsmath}
\usepackage{amssymb}
\usepackage{textcomp}
\usepackage{graphicx}
\usepackage{color}
\usepackage{verbatim}
\usepackage{times}
\usepackage{mathpazo}

\newcommand{\etal}{{\em et al. }}

\begin{document}

\title{Surface-induced First Order Transition in Athermal
Polymer/Nanoparticle Blends}

\author{E. S. McGarrity}
\affiliation{Department of Chemical Engineering \& Materials Science, Michigan
State University, East Lansing, Michigan, 48824-1226, USA}
\author{A. L. Frischknecht}
\affiliation{Sandia National Laboratories, Albuquerque, New Mexico 87185, USA}
\author{L. J. D. Frink}
\affiliation{Sandia National Laboratories, Albuquerque, New Mexico 87185, USA}
\author{M. E. Mackay}
\affiliation{Department of Chemical Engineering \& Materials Science, Michigan
State University, East Lansing, Michigan, 48824-1226, USA}

\begin{abstract}
We investigate the phase behavior of athermal polymer/nanoparticle blends near a hard substrate. We apply the density functional theory of Tripathi and Chapman to these blends. We find a first order phase transition where the nanoparticles expel the polymer from the surface to form a monolayer. The transition density depends on the polymer length and the system bulk density. The effect is due to the packing entropy of the species and  configurational entropy of the polymer. The simplicity of the system allows us to understand the so-called ``entropic-push'' observed in experiments. 
\end{abstract}


\maketitle

Mixtures of polymers and colloidal-size particles have been shown to exhibit
rich phase behavior, both in bulk and near surfaces \cite{Poon2002}.
The corresponding behavior of blends of nanosize particles and polymers
is still being explored and is not well understood.  The small size of the
nanoparticles leads to both new phenomena \cite{Tuteja2005,mackay-2006,Hooper2006} and
potential new applications.  Nanoparticles have been shown to migrate
to interfaces such as cracks or substrates in experiments, and thus may
allow for self-healing surfaces \cite{balazs-2005-1}.  The self-assembly of
nanoparticles in, for example, diblock copolymers has been studied with the
goal of producing composites with useful mechanical, magnetic, and optical
properties \cite{Thomas2003,Lin2005}.

However, a detailed understanding of the mechanisms by which nanoparticles interact with a polymer matrix near a substrate is lacking.  In general, such systems involve a complex admixture of entropic, enthalpic, and chemical interactions between the different components that is difficult to ferret out by experiment.  Insight into the physics of polymer nanocomposites can be gained by computational studies of simplified models in order to isolate the individual interactions.   

Additionally, interesting phenomena occur even in relatively simple polymer nanocomposites.  Recent neutron reflectivity experiments \cite{krishnan-2005,krishnan-2007} have found that in ultra-thin films of polymer/nanoparticle mixtures, the nanoparticles often form a monolayer on the substrate.  Remarkably, this occurs for the case of polystyrene blended with polystyrene nanoparticles \cite{krishnan-2005}, in a regime where the nanoparticles are miscible in the polymer in the bulk \cite{mackay-2006}.  Since the polymer and nanoparticles are chemically compatible, the forces driving the nanoparticles to the substrate are thought to be entropic in origin \cite{krishnan-2007,krishnan-2007-2}.  The presence of about a monolayer of nanoparticles at the substrate has the additional surprising property of preventing the dewetting of the films. 

In this Letter we provide theoretical evidence in support of the hypothesis that purely entropic driving forces can result in nanoparticles segregating to the substrate when blended with a polymer melt.   We explore the behavior of the most simple relevant system: a mixture of hard nanoparticles in a hard-chain polymer melt, near a hard surface.  An understanding of this system is a prerequisite for understanding more complex situations in polymer nanocomposite films (such as the effect of attractive interactions) since the entropic effects will always be present.  As we will show, we find a previously unknown surface phase transition in which the nanoparticles do indeed spontaneously form a monolayer on the substrate.  

We employ a computationally efficient, classical density functional theory
(DFT) to explore the phase behavior of our model system.  DFTs can capture
the microscopic structure and thermodynamic behaviors of complex fluids
\cite{wu-2006,wu-2007}. They are based on minimization of a grand potential 
free energy functional.  Since the minimization results in the grand potential free 
energy of the system, DFTs are well-suited to the study of phase behavior.
Previous theoretical studies of polymer/nanoparticle mixtures have mostly focused 
on block copolymers and used various techniques.
Recently Sides {\em et al}.\ used a hybrid particle-field theory (HPF) \cite{sides-2006}
in which the nanoparticles are treated as an external field and the polymer
is described with self-consistent field theory (SCFT).  Earlier work used a combined SCFT/DFT method \cite{ginzburg-2002}, with the polymers described by SCFT and the nanoparticles by a DFT. 
Other works that treated blends with DFT include an examination of the wetting behavior of mixtures of colloids and polymers near the bulk coexistence \cite{bryk-2005-3}, and an investigation of nanoparticles in diblock copolymer thin films \cite{Cao2007}.  
Here we also treat both the nanoparticles and
the polymer within the same, consistent theoretical framework.  Our work is the first to examine particles in homopolymer melts near a substrate with DFT.

We use a density functional that is based on the fundamental measure theory
of hard sphere liquids, pioneered by Rosenfeld \cite{rosenfeld-1989}.
Both the nanoparticles and the polymer segments are treated as hard spheres.
Bonding constraints between the polymer segments are enforced using the
Wertheim-Tripathi-Chapman (WTC) bonding functional \cite{tripathi-2005-1,
tripathi-2005-2}, which is based on Wertheim's thermodynamic perturbation
theory (TPT1) \cite{wertheim-1984-1,wertheim-1984-2}.  The DFT is
formulated in an open ($\mu$VT) ensemble.  The grand potential for the hard
sphere/polymer blend is \begin{align}
  \Omega\left[ \rho_{\alpha}({\bf r})\right] &=
            F_{id}\left[  \rho_{\alpha}({\bf r}) \right]
           +  F_{hs}\left[  \rho_{\alpha}({\bf r}) \right] + F_{ch}\left[
           \rho_{\alpha}({\bf r}) \right] \\ \notag
          & + \sum_{\alpha} \int d {\bf r} \rho_{\alpha}({\bf r})
               \left[V_{\alpha}({\bf r}) - \mu_{\alpha}\right] ,
\end{align} where the terms on the right-hand side represent the Helmholtz
free energies for the ideal gas, the hard sphere, and the chain constraints.
The final term is the Legendre transformation where the $\mu_{\alpha}$ are
the site chemical potentials and $V_{\alpha}({\bf r})$ is an external field.
The exact form of the Rosenfeld hard sphere term, $F_{hs},$ can be found in \cite{rosenfeld-1997}. A more accessible derivation of this functional can be found in \cite{roth-2002}.  We implemented a form of the chain free energy functional $F_{ch}$ (Eq.\ 22 from \cite{tripathi-2005-1}) to keep track of,
and solve explicitly for, segment densities by treating each segment as a
separate species.  The Rosenfeld functional that we use is known to give very
accurate density profiles for binary mixtures of hard spheres \cite{roth-2000}, while the WTC
functional has been shown to accurately capture the physics of homopolymers
and polymer blends near surfaces \cite{tripathi-2005-1,tripathi-2005-2}.

Minimization of $ \Omega\left[ \rho_{\alpha}({\bf r})\right]$  leads to a set of nonlinear
integral equations for the density distributions, $\rho_{\alpha}(\bf{r}),$
of the constituent species.  We solve the DFT equations using the Tramonto
fluids DFT code \cite{frink-2000-1,frink-2000-2,tramonto}, with numerical
methods as detailed elsewhere \cite{heroux-2007}.  
The polymers consist of a chain of $N$ freely-jointed tangent spheres
with diameter $\sigma_{p}$ while the nanoparticles are hard spheres with
diameter $\sigma_{n}=2\sigma_p$.  We keep the total  packing fraction fixed at
$\eta=0.3665$ unless otherwise noted, where $\eta = \pi/6(\rho_n\sigma_n^3 +
\rho_p\sigma_p^3)$, and $\rho_{\alpha}$ and $\sigma_{\alpha}$ are the species
bulk density and diameter, respectively.   The density
profiles were assumed to vary in the $z$ direction only.  We performed our calculations
in a large ($L=80 \sigma_p$) box with reflective boundary conditions to guarantee bulk behavior in the middle of the box.  The external field representing the hard wall is  $V_{\alpha} = \infty$ for $z/\sigma_p<\sigma_{\alpha}$ and  $V_{\alpha}
= 0$ for $z/\sigma_p>\sigma_{\alpha}$.

We calculated the free energy and density profiles of the polymer/nanoparticle
blend as a function of nanoparticle concentration.  We began by
converging a solution at a nanoparticle density of $\rho_n \sigma_p^3
=0.001$. We then employed the arc-length continuation algorithms found in the LOCA software package
\cite{keller-1977, salinger-2005} to trace the free energy of the system as a function of 
nanoparticle density, at constant packing fraction
$\eta$. This allowed us to find regions of multiple solutions, with radically
different morphologies.

\begin{figure}[htbp]
\begin{center}
\includegraphics[width=0.75\columnwidth]{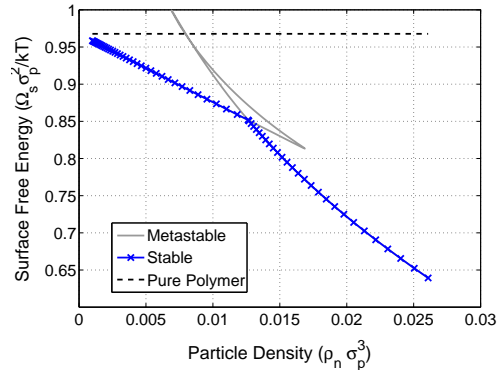}
\caption{Surface free energy versus nanoparticle density for
$N=40$ and $\eta=0.3665$. The dark curve with the ``x's'' indicates stable
configurations. The abrupt change in slope of this curve at $\rho_n^*=0.01263$
is indicative of a first order phase transition. Solutions in the metastable and unstable regions are
shown by the light curve. The
dashed line indicates the free energy of a neat polymer system with $N=40$
and $\eta=0.3665$. \label{fig:Omega_s}}
\end{center}
\end{figure}

The surface free energy of a blend with $N=40$ is shown in Fig.\
\ref{fig:Omega_s}, where the surface free energy is defined as $\Omega_s
\left[ \rho_{\alpha}(\bf{r}) \right] =\Omega \left[ \rho_{\alpha}(\bf{r})
\right] -\Omega_{bulk}$, and $\Omega_{bulk}$ is the free energy of a
homogeneous bulk system with the same packing fraction and composition.  The dark curve with
``x'' markers corresponds to the stable solutions of the free energy
minimization for the mixture. There is a distinct change in slope in the curve at a
density of $\rho^*_n\sigma_p^3=0.01263$, which indicates a first order phase
transition. The light part of the curve represents the metastable and unstable
branches of the phase space explored by the continuation solver. The point at
which the curve crosses itself has two solutions with distinct morphologies.
The dashed line represents the free energy of a neat polymer system with
the same length ($N=40$) and packing fraction.   Note that the addition 
of nanoparticles reduces the free energy of the system.

\begin{figure}[h!]
\begin{center}
\includegraphics[width=0.75\columnwidth]{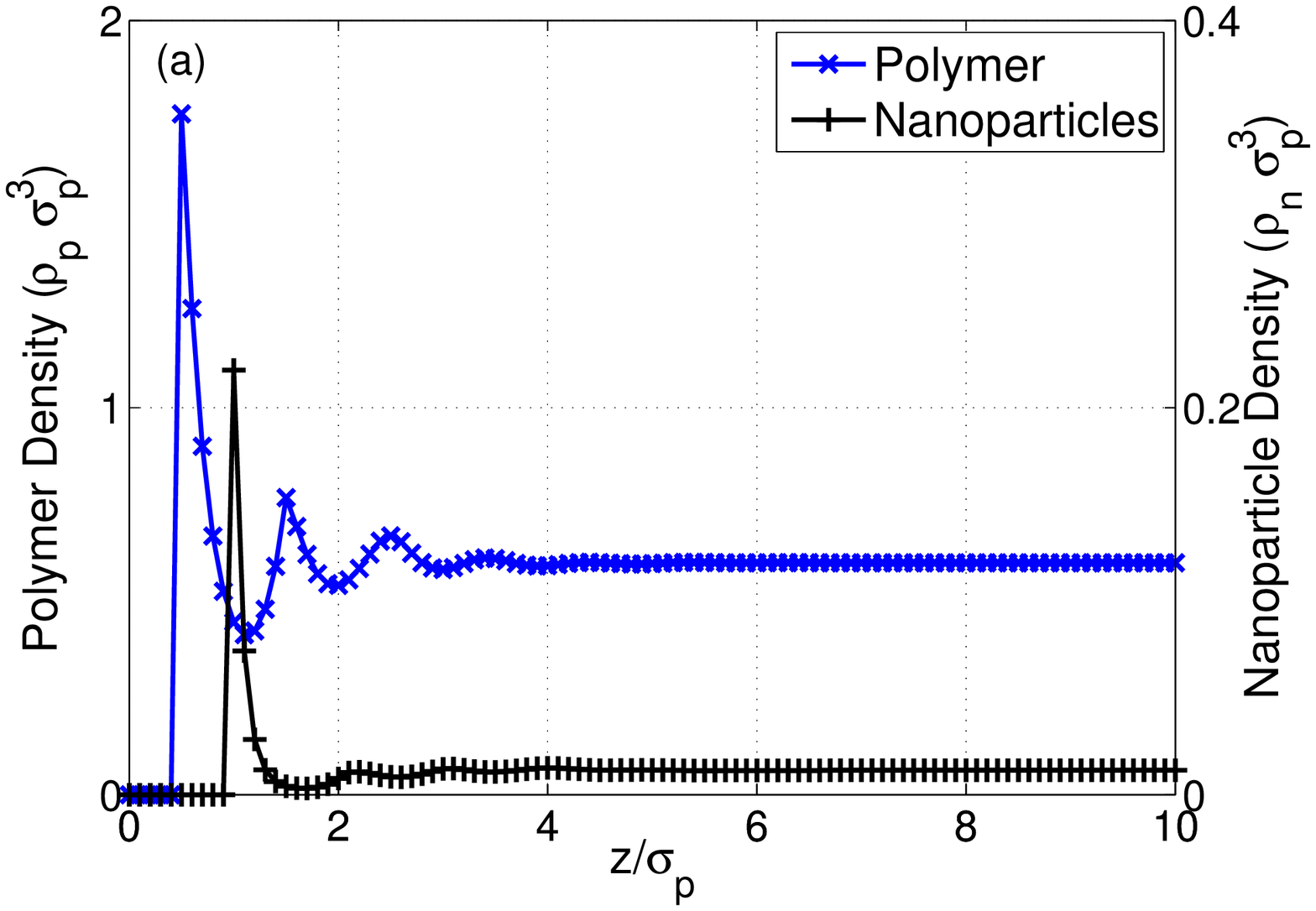}
\includegraphics[width=0.75\columnwidth]{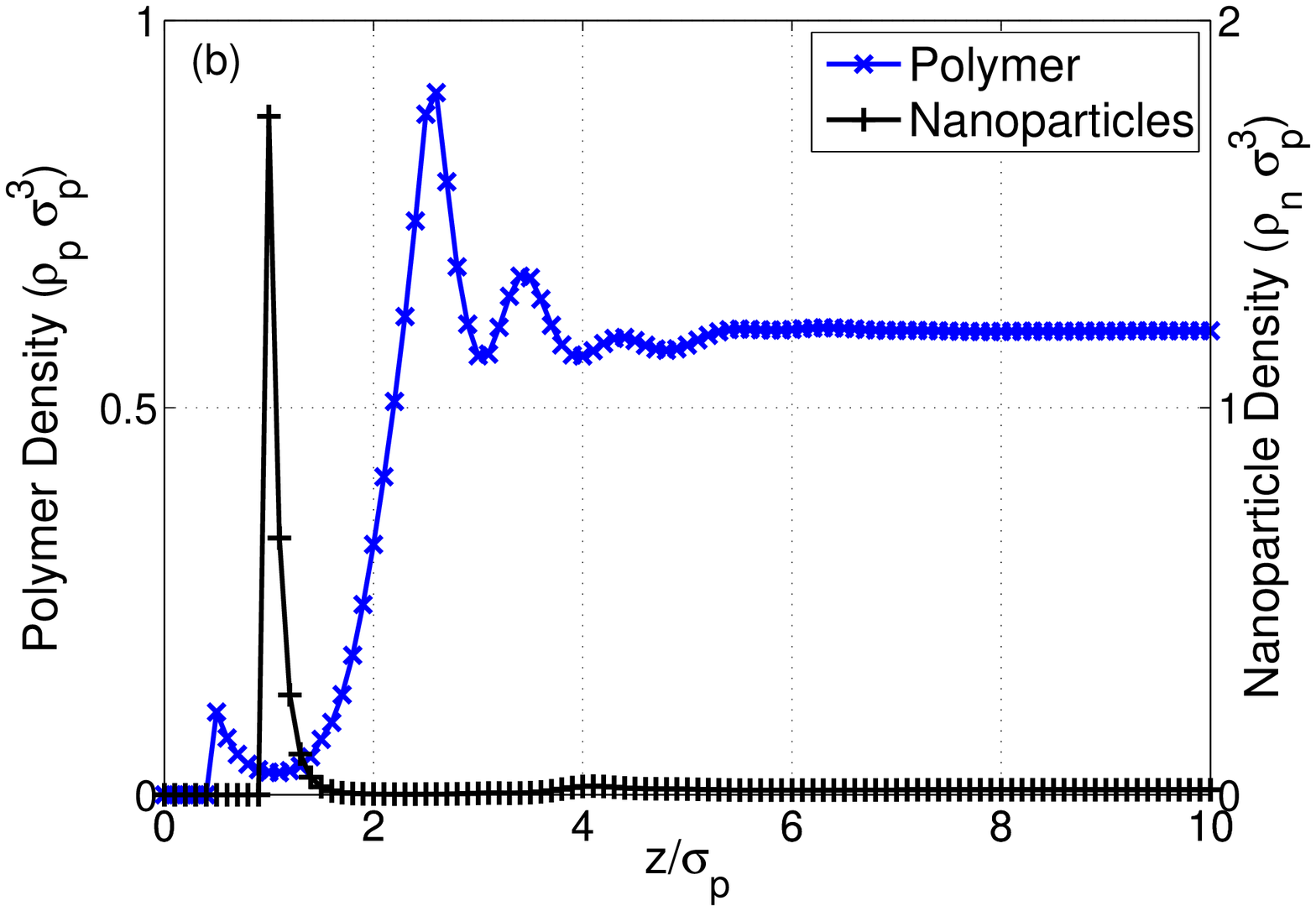}
\caption{Density profiles for polymer (``x's'' -- left scale) and nanoparticles (``+'s'' -- right scale),
with $N=40$ with $\eta=0.3665$ at
the phase transition ($\rho_n^* = 0.01263$). These profiles were computed by
setting the density to $\rho_n^*$ and restarting with an initial guess profile which was
converged at a lower (a) and higher (b) density.
  \label{fig:dens1}}
\end{center} 
\end{figure}

The two coexisting density profiles found at the phase transition in Fig.\
\ref{fig:Omega_s}  are shown in Fig.\ \ref{fig:dens1}.  
A density profile converged at the coexistence
density ($\rho^*_n$) from a profile at a slightly lower nanoparticle density is shown in Fig.\ 
\ref{fig:dens1}a.  The density
profile is typical for a dense liquid.  We see that both the nanoparticles
and polymer have pronounced peaks near the substrate.  The polymer peak is closer
to the substrate because of the smaller size of the polymer segments.
Converging to $\rho^*_n$ from a higher particle density (see Fig.\
\ref{fig:dens1}b), we find that the polymer has
been almost completely excluded from the vicinity of the substrate, and
there is a large peak in the nanoparticle density adjacent to the substrate, indicating a large adsorption of nanoparticles.  The height of the first peak in the nanoparticle
density does not change significantly as we add nanoparticles above the transition density,
so the structure of the monolayer remains the same above the transition. The contact densities are in reasonable agreement with the pressure sum rule $p/kT = \sum_{\alpha} \rho_{\alpha}(z=\sigma_{\alpha}/2+\sigma_p/2)$.

\begin{figure}[h!]
\begin{center}
\includegraphics[width=0.75\columnwidth]{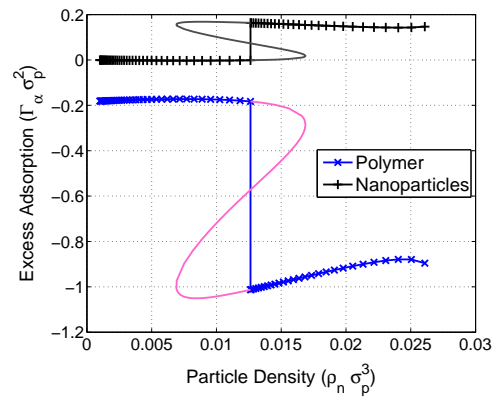}
\caption{Excess adsorption of blend components as a function of nanoparticle
density. The ``x's''  and ``+'s''  indicate  the polymer and nanoparticles, respectively. 
The light colored parts of the curves correspond to the meta/unstable branches. 
\label{fig:ads_ex}}
\end{center}
\end{figure}

The excess adsorption $\Gamma_{\alpha}$ is shown in Fig.\ \ref{fig:ads_ex},
where $\Gamma_{\alpha}$ is defined as 
$\Gamma_{\alpha} = \int_0^\infty dz \left(\rho_{\alpha}(z)-\rho_{\alpha}\right)$.
The vertical lines indicate the sudden jump in adsorption as the nanoparticle
density is increased through the phase transition. This jump shows the
polymer being expelled from the wall by the nanoparticles with the van der
Waals loops being clear indicators of a first order transition.

Thus we have found a first order phase transition in which the
polymer is pushed away from the substrate and is replaced by the
nanoparticles. Integrating under the first peak $(0 \leq z/\sigma_p \leq 2)$ of the
nanoparticle density profile in Fig.\ \ref{fig:dens1}b gives a density per
unit area of 0.2866/$\sigma_p^2$, which corresponds to a densely packed
monolayer on the surface with an areal coverage of 0.9.

To our knowledge this is the first report of the calculation of
an entropically-driven surface phase transition in an athermal
polymer/nanoparticle blend.  We note that unlike other surface phase
transitions in similar systems \cite{ayadim-2006,bryk-2005-2, bryk-2006},
our blend does not sit near a corresponding bulk phase transition.
Previous studies of binary hard sphere blends have found that there is a
fluid-fluid demixing transition only when the size ratio of the particles
exceeds $5:1$ \cite{rosenfeld-1997}.   Paricaud \etal \cite{paricaud-2003} showed (for a blend
with a similar equation of state to ours) that there is also a first order
fluid-fluid demixing transition in hard polymer/particle mixtures, but only
when the particle diameter is larger than about 5 times the monomer diameter.  Similar immiscibility for large size asymmetries is
predicted by Hooper and Schweizer \cite{Hooper2004}.  Our system should thus
be in the regime of bulk miscibility.  

\begin{figure}[h]
\begin{center}
\includegraphics[width=0.75\columnwidth]{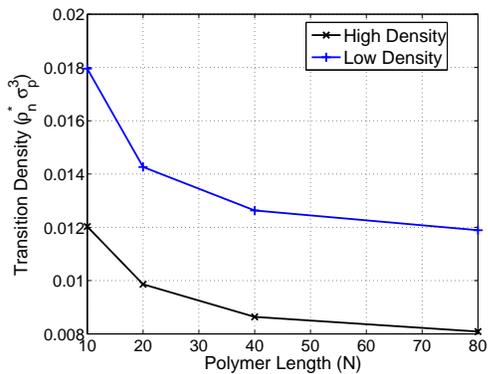}
\caption{Nanoparticle density at the phase transition as a function of chain length. The
``+'s'' and ``x's'' are for $\eta=0.3665$ and $\eta= 0.4152$, respectively.
  \label{fig:LengthDep}}
\end{center}
\end{figure}

In Fig.\ \ref{fig:LengthDep} we show the effects of changing the length
of the polymer on the density of the particles  $\rho_n^*$ at the transition. Results are shown
for two systems, one with a solution-like packing fraction ($\eta=0.3665$)
and the second with a melt-like packing fraction ($\eta=0.4152$). In both
cases, $\rho_n^*$ decreases with the length of the polymer.
For sufficiently short chains, ($N \leq 8$ for $\eta=0.3665$ and $N \leq 5$ for
$\eta=0.4152$) we find no phase transition at all.  Also, there is no
phase transition for a binary mixture of hard spheres of diameters $\sigma_p$ and
$2 \sigma_p$.  Thus, the phase transition is a polymeric effect.

The transition must be due to the interplay of chain entropy and packing
entropy.  One driving force is the well-known depletion potential of a
large sphere immersed in a fluid of smaller spheres \cite{dijkstra-1999}.  This effect leads to
enrichment of large spheres near a substrate.  Packing entropy should thus help to
drive the nanoparticles to the substrate.  We note that for particles
with the same size as the monomer, $\sigma_n=\sigma_p$,
a phase transition in not found, so the particle/monomer size asymmetry is important.
A second driving force is the loss of conformational entropy  when the
polymer is close to the wall.  Apparently, the conformational entropy loss
is key since the transition only happens for sufficiently long polymers.


To conclude, we have identified a surface-induced first order phase
transition in athermal polymer/nanoparticle blends. This transition is governed by the polymer
chain length and nanoparticle concentration. The existence of a
phase transition in such a simple system is noteworthy because the entropic contributions at play here will also be factors in more complex attractive systems. 
Our work takes a step towards identifying and clarifying the effects of the subtle interplay among entropic contributions originating 
from size anisotropy and from bonding constraints in polymer/nanoparticle blends.

The existence of this surface phase transition is consistent with the observation of a monolayer of nanoparticles at the substrate in the experiments mentioned previously.  We have demonstrated that entropic driving forces alone are sufficient to form a monolayer of nanoparticles at a substrate, adding weight to the ``entropic-push'' hypothesis used to describe the experimental results \cite{krishnan-2007,krishnan-2007-2}.   In addition, the theory predicts a lowered surface free energy induced by
adding nanoparticles to a supported polymer film.  The lowered surface free energy may contribute to the observed inhibition of dewetting caused by the addition of nanoparticles \cite{krishnan-2005}.  

\vspace*{1.5mm}
{\bf Acknowledgments: }We thank Phil Duxbury and Frank van Swol for helpful
discussions. Thanks also to the U.S. Department of Energy for funding
this research (contract DE-FG02-05ER46211). This work was performed in part
at the US Department of Energy, Center for Integrated Nanotechnologies,
at Los Alamos National Laboratory (Contract DE-AC52-06NA25396) and Sandia
National Laboratories (Contract DE-AC04-94AL85000).


\end{document}